\documentclass{aa}  

\usepackage{graphicx}
\usepackage{txfonts}
\usepackage{color}
\usepackage{stfloats}
\usepackage{multirow}
\usepackage{mhchem}
\usepackage{booktabs}
\usepackage{hyperref}

\newcommand{\acsesc}{ACS Earth Space Chem.}

\newcommand{\cpl}{Chem. Phys. Lett.}

\newcommand{\gcacta}{Geochim. Cosmochim. Acta}

\newcommand{\jcsft}{J. Chem. Soc. Faraday Trans.}

\newcommand{\jms}{J. Mol. Spectr.}
\newcommand{\jmst}{J. Mol. Struct.}
\newcommand{\joc}{J. Org. Chem.}

\newcommand{\nature}{Nature}
\newcommand{\natastro}{Nat. Astron.}
\newcommand{\oex}{Opt. Express}
\newcommand{\ogc}{Org. Geochem.}

\newcommand{\pnas}{PNAS}
\newcommand{\rpp}{Rep. Prog. Phys.}

\newcommand{\rsi}{Rev. Sci. Instrum.}
\newcommand{\science}{Science}

\begin{document}

\title{D/H and $^{12}$C/$^{13}$C ratios of benzonitrile and indene in TMC-1\\ support bottom-up formation of polycyclic aromatic hydrocarbon-like molecules\thanks{Based on observations carried out with the Yebes 40m telescope (projects 19A003, 20A014, 20D023, 21A011, 21D005, and 23A024). The 40m radio telescope at Yebes Observatory is operated by the Spanish Geographic Institute (IGN; Ministerio de Transportes, Movilidad y Agenda Urbana).}}

\titlerunning{D/H and $^{12}$C/$^{13}$C ratios of benzonitrile and indene in TMC-1}
\authorrunning{Heras et al.}

\author{
D.~Heras\inst{1},
R.~Mora-L\'opez\inst{2,3},
J.~Janeiro\inst{4},
M.~N.~Alparslan\inst{2,5},
C.~Cabezas\inst{2},
M.~Ag\'undez\inst{2},
D.~P\'erez\inst{4},
I.~Peña\inst{1},
M. Pereira-Santaella\inst{2},
M.~A.~Crutchfield\inst{1},
J.~R.~Mor\'an\inst{1},
A.~L.~Steber\inst{1},
A.~Lesarri\inst{1},
C.~P\'erez\inst{4,6},
I.~Garc\'ia-Bernete\inst{7},
N.~Marcelino\inst{8,9},
B.~Tercero\inst{8,9},
J.~R.~Goicoechea\inst{2},
and
J.~Cernicharo\inst{2}
}

\institute{
Departamento de Qu\'imica Física y Qu\'imica Inorg\'anica, Facultad de Ciencias-I.U. CINQUIMA, Universidad de Valladolid, 47011 Valladolid, Spain
\and
Departamento de Astrof\'{i}sica Molecular, Instituto de F\'{i}sica Fundamental, CSIC, Calle Serrano 123, 28006 Madrid, Spain\\\email{carlos.cabezas@csic.es, marcelino.agundez@csic.es}
\and
Universidad de Alcal\'a, Departamento de F\'isica y Matem\'aticas, Campus Universitario, 28871 Alcal\'a de Henares, Madrid, Spain
\and
Centro Singular de Investigaci\'on en Qu\'imica Biol\'oxica e Materiais Moleculares (CiQUS) and Departamento de Qu\'imica Org\'anica, Universidade de Santiago de Compostela, 15782 Santiago de Compostela, Spain\\
\email{dolores.perez@usc.es}
\and
Department of Molecular Biology and Genetics, Faculty of Science and Letters, Istanbul Technical University, Istanbul 34469, Turkey
\and
Oportunius. Axengia Galega de Innovaci\'on, 15702 Santiago de Compostela, Spain
\and
Centro de Astrobiolog\'ia (CAB), CSIC-INTA, Camino Bajo del Castillo s/n, 28692 Villanueva de la Ca\~nada, Madrid, Spain
\and
Observatorio Astron\'omico Nacional, IGN, Calle Alfonso XII 3, 28014 Madrid, Spain
\and
Observatorio de Yebes, IGN, Cerro de la Palera s/n, 19141 Yebes, Guadalajara, Spain
}

\date{Received; accepted}

\abstract
{Polycyclic aromatic hydrocarbons (PAHs) have recently been detected in cold dark clouds through their rotational spectra, but it is currently a matter of debate whether they form through a bottom-up process from small hydrocarbons or through a top-down scenario involving the fragmentation of large PAHs inherited from a previous evolutionary stage of diffuse cloud. Isotopic ratios can shed light on the origin of PAHs in cold dark clouds. Large aromatic molecules observed in the infrared are enriched in deuterium and may also be enriched in $^{13}$C because they likely formed in the ejecta of evolved stars. We synthesized indene deuterated at the aliphatic position in the laboratory and measured its rotational spectrum, together with those of the nine $^{13}$C singly substituted isotopomers of indene. We used the measured spectra of the D and $^{13}$C isotopic species of indene, together with previously measured spectra of D and $^{13}$C isotopic benzonitrile, to search for these species in the cold dark cloud \mbox{TMC-1} using QUIJOTE data obtained with the Yebes\,40m telescope. We used a line stack method to tentatively detect deuterated benzonitrile and detect $^{13}$C-substituted benzonitrile, but find no evidence of either the D or $^{13}$C isotopic species of indene. We derive an aromatic D/H ratio of 0.53\,\% from benzonitrile, an aliphatic D/H ratio $<$\,3.3\,\% from indene, and $^{12}$C/$^{13}$C ratios of 126 and $>$\,55 for benzonitrile and indene, respectively. These values are consistent with isotopic ratios derived for other molecules in \mbox{TMC-1} and show no enrichment in either D or $^{13}$C that could be ascribed to inheritance from large PAHs in a top-down scenario; a bottom-up scenario is therefore the most likely.}

\keywords{astrochemistry -- line: identification -- ISM: individual objects (\mbox{TMC-1}) -- ISM: molecules -- radio lines: ISM}

\maketitle

\section{Introduction}

Polycyclic aromatic hydrocarbons (PAHs) have long been suspected to be ubiquitous and abundant in the interstellar medium as carriers of aromatic infrared bands (AIBs); these bands are observed in almost every astronomical environment exposed to ultraviolet (UV) radiation \citep{Leger1984,Allamandola1985,Tielens2008}. Due to the nonspecific nature of the infrared bands, there is still a debate about whether PAHs or other large organic molecules are carriers of AIBs \citep{Kwok2022,Tielens2026}. Unambiguous identification of PAHs in the interstellar medium was achieved only a few years ago, with the detection through rotational spectroscopy of individual molecules in, paradoxically, the UV-shielded environment of cold dark clouds. The current inventory of PAH-like molecules detected in the emblematic starless core \mbox{TMC-1} comprises single rings (cyclopentadiene and benzene), two-fused rings (indene and naphthalene), three-fused rings (acenaphthylene, cyclopentindene, and phenalene), the four-ring species pyrene, and the seven-ringed molecule coronene \citep{Cernicharo2021a,Cernicharo2021b,Cernicharo2024a,Cernicharo2026,McGuire2018,McGuire2021,McCarthy2021,Lee2021,Burkhardt2021a,Wenzel2024,Wenzel2025a,Wenzel2025b,Cabezas2025,Fuentetaja2026}. For most of these cycles, observations have detected the polar cyano derivative rather than the pure hydrocarbon. In addition to \mbox{TMC-1}, some of these cycles have also been observed in other cold dark clouds \citep{Burkhardt2021b,Agundez2023a,Wenzel2026} and in warm molecular clouds in the Galactic center \citep{Rivilla2026}. Polycyclic aromatic hydrocarbons have also been unequivocally identified in Solar System bodies such as the Murchison meteorite \citep{Basile1984} and the Ryugu asteroid \citep{Zeichner2023}.

Polycyclic aromatic hydrocarbon-like molecules are clearly present in cold dark clouds with large abundances (10$^{-9}$-10$^{-8}$ relative to H$_2$; \citealt{Agundez2026}), but it is not clear how they form in this particular kind of environment. Evidence based on the spatial similarity of carbon chains and cycles \citep{Cernicharo2023} and the correlation in the abundance of these two types of molecules \citep{Agundez2023a} points to a bottom-up scenario in which PAH-like species form from small hydrocarbon molecules in situ in the cold dark cloud. However, there is a debate on which chemical pathways are sufficiently efficient to account for the large abundances observed \citep{Yang2024,Mallo2025,Jacovella2026,Loison2026,Agundez2026}. In addition, the abundances of PAH-like species in \mbox{TMC-1} are rather uniform and independent of size \citep{Wenzel2025a,Wenzel2025b,Agundez2026}, which is difficult to explain within the framework of a bottom-up scenario. An alternative is a top-down scenario in which the observed PAH-like molecules form through the fragmentation of large PAHs (those responsible for the AIBs) inherited from the previous evolutionary stage of diffuse cloud \citep{Pety2005,Zhen2014,Burkhardt2021b,Goicoechea2025,Garcia-Bernete2026}.

Isotopic ratios can be used to discern between the bottom-up and top-down scenarios. For example, observations of AIBs at 4.65 $\mu$m and 4.35 $\mu$m in Galactic UV-illuminated regions and external galaxies indicate that AIBs carriers are enriched in deuterium \citep{Peeters2004,Peeters2024,Onaka2014,Doney2016,Boersma2023,Pereira-Santaella2024,Yang2025,Draine2025}. Infrared observations do not provide constraints on the $^{12}$C/$^{13}$C ratio of AIB carriers, but the origin of these large aromatic molecules is commonly ascribed to the ejecta of evolved stars \citep{Kwok2004,Tielens2008}, which are enriched in $^{13}$C compared with the interstellar medium \citep{Milam2009,Ramstedt2014,Ziurys2020}. Therefore, if the PAHs unambiguously identified in cold dark clouds form through fragmentation of large PAHs inherited from the previous diffuse cloud stage in a top-down scenario, we would expect them to be enriched in deuterium and $^{13}$C.

In \cite{Steber2025} we report the rotational spectrum of deuterated benzonitrile and an unsuccessful search for it in \mbox{TMC-1}. We derived a  D/H ratio of $<$\,1.2\,\%, which argued against the top-down origin of PAHs in cold dark clouds. Here we report a more sensitive search for deuterated and $^{13}$C-substituted benzonitrile and indene using a line stack analysis to discern between the bottom-up and top-down scenarios. We took rotational spectroscopic data from \cite{Steber2025} for deuterated benzonitrile, from archive data for $^{13}$C benzonitrile, and from measurements carried out in this work for deuterated and $^{13}$C indene, for which the deuterated form was synthesized in the laboratory. In Sect.\,\ref{sec:spectroscopy} we present the rotational spectroscopy of all these species, in Sect.\,\ref{sec:search} we describe the astronomical search for them in \mbox{TMC-1} and the derived abundances or upper limits, in Sect.\,\ref{sec:discussion} we discuss the implications for the origin of PAHs in cold dense clouds, and in Sect.\,\ref{sec:conclusions} we present our conclusions.

\section{Rotational spectroscopy of D and $^{13}$C benzonitrile and indene} \label{sec:spectroscopy}

\begin{figure}
\centering
\includegraphics[angle=0,width=\columnwidth]{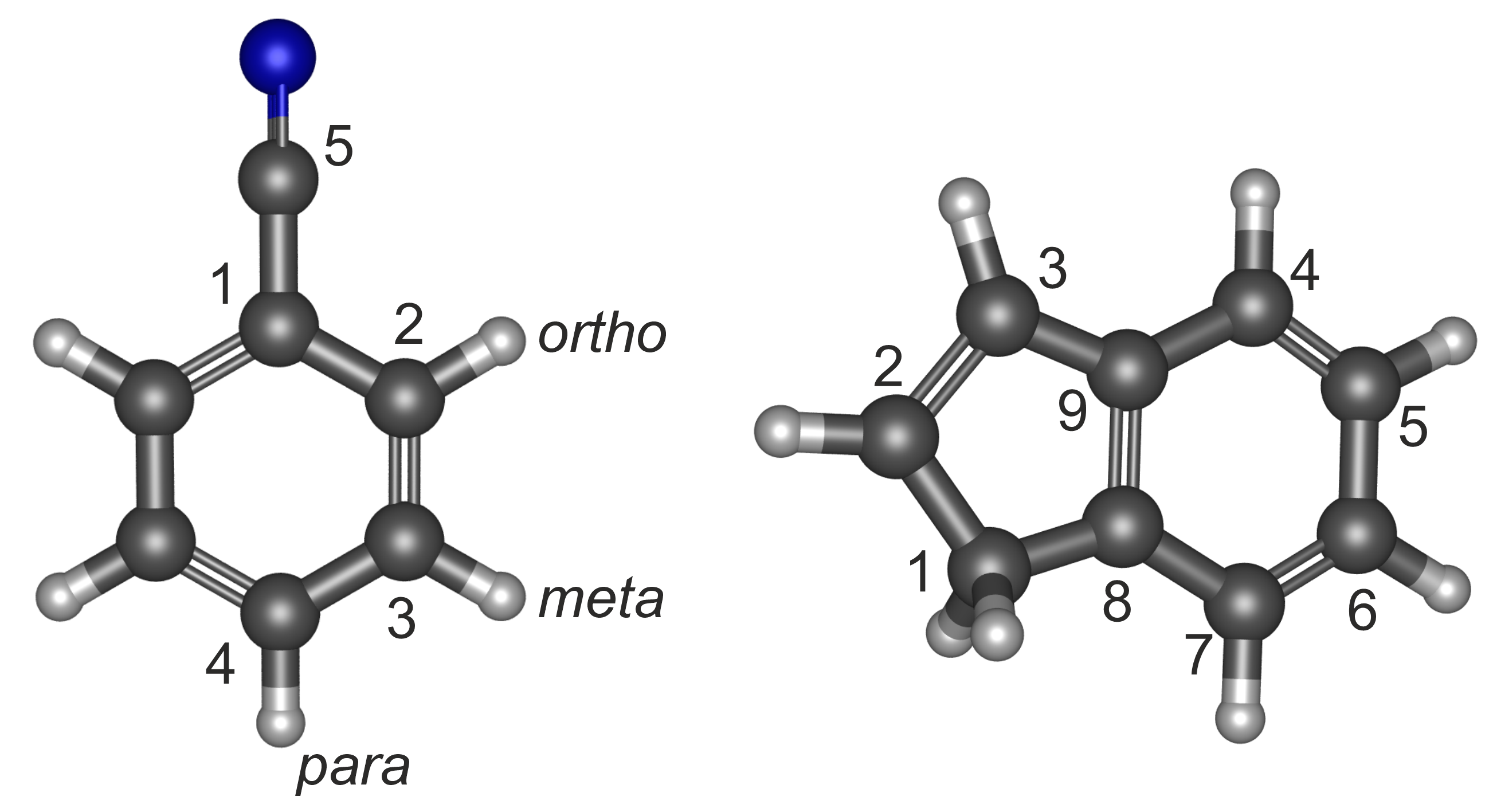}
\caption{Molecular structures of benzonitrile (left) and indene (right). The numbers indicate the carbon atom labels used for the $^{13}$C isotopic species and the ortho, meta, and para positions used for the D isotopic species of benzonitrile.} \label{fig:molecules}
\end{figure}

We took the rotational spectroscopy of deuterated benzonitrile from \cite{Steber2025}, while for $^{13}$C benzonitrile, we analyzed line frequencies reported in the literature to derive spectroscopic parameters and predict line frequencies. For D- and $^{13}$C indene we performed laboratory spectroscopic measurements, which included three different experiments. For deuterated indene, we observed the spectrum of the isotopomer substituted at carbon position 1 (see Fig.\,\ref{fig:molecules}) in the microwave (2-8 GHz) and millimeter domain (75-110 GHz) ranges. For these measurements, we used a synthesized sample of 1d-indene. Indene has six other deuterated isotopomers, but here we report only the spectroscopy of the deuterated form at carbon position 1, which is the only carbon with an aliphatic, that is,  nonaromatic, character. This isotopomer is of greater astronomical interest because observations of AIBs suggest that deuteration of PAHs in UV-illuminated regions preferentially occurs at aliphatic rather than aromatic carbon sites (see Sect.\,\ref{sec:discussion}). For the $^{13}$C isotopologue of indene, we observed the spectra of the nine $^{13}$C isotopomers (see Fig.\,\ref{fig:molecules}) in natural abundance in the 2-18 GHz range using a commercial sample of indene. The details of the spectral analysis of $^{13}$C benzonitrile, the chemical synthesis of 1d-indene, the spectroscopic measurements of D and $^{13}$C indene, and the analysis of the corresponding spectra are given below.

\subsection{Rotational transition frequencies of $^{13}$C benzonitrile}

The rotational spectra of benzonitrile isotopologues were first studied by \cite{Bak1962} in the 12-25 GHz frequency range, with a frequency accuracy of around 500 kHz. Later, \cite{Casado1971} analyzed the microwave spectra of nine isotopologues up to 26 GHz and determined precise rotational constants using measurements with accuracies of 50-100 kHz. A third investigation was reported by \cite{Dahmen1994}, who provided accurate rotational constants derived using Fourier transform microwave (FTMW) spectroscopy in the 6-18 GHz frequency range. We obtained accurate predictions for the rotational transition frequencies of the $^{13}$C species of benzonitrile in the Q band by combining the data reported by \cite{Casado1971} and \cite{Dahmen1994}. The merged fit for each isotopic species includes all the lines measured by \cite{Dahmen1994} up to 18 GHz and the frequencies reported by \cite{Casado1971} above 18 GHz. We proceeded in this manner because the frequency accuracy of the measurements of \citet{Dahmen1994} is 2 kHz, whereas the measurements reported by \cite{Casado1971} are less accurate. We analyzed the data using the SPFIT program \citep{Pickett1991} with the $A$-reduction of the Watson's Hamiltonian in the $I^r$ representation \citep{Watson1977}. The analysis allowed us to determine the values of the rotational constants, the $^{14}$N nuclear quadrupole coupling constants, and the centrifugal distortion constants $\Delta_J$, $\Delta_{JK}$, and $\delta_J$, while $\Delta_{K}$ and $\delta_K$ were fixed to the values reported by \citet{Zdanovskaia2018}. The spectroscopic parameters obtained from our fits, which are given in Table\,\ref{table:13C-benzonitrile}, allowed us to predict the rotational transition frequencies of the $^{13}$C isotopic species of benzonitrile in the Q band with uncertainties of less than 30 kHz for $a$-type transitions with $K_{a}$ values $<$\,4.

\begin{table*}
\small
\caption{Spectroscopic parameters of $^{13}$C isotopic species of benzonitrile.}
\label{table:13C-benzonitrile}
\centering
\begin{tabular}{lccccc}
\hline
Parameter                &  $^{13}$C$_1$  & $^{13}$C$_2$  & $^{13}$C$_3$  & $^{13}$C$_4$  & $^{13}$C$_5$  \\
\hline
$A$/MHz                  &     5655.5140(113)   &     5563.9117(58)   &     5565.6502(103)  &    5655.4926(219)   &  5655.2551(282)    \\
$B$/MHz                  &    1545.551877(185)  &   1546.803376(107)  &   1535.713007(179)  &   1523.65625(70)    &  1528.64095(33)    \\
$C$/MHz                  &    1213.601513(111)  &   1210.089832(74)   &   1203.373135(126)  &   1200.05748(32)    & 1203.136701(272)   \\
$\Delta_J$/kHz           &      0.04629(240)    &       0.04464(113)  &       0.04333(184)  &       0.0566(71)    &    0.0475(38)      \\
$\Delta_{JK}$/kHz        &       0.886(47)      &        0.962(65)    &         1.012(107)  &        0.648(155)   &    0.850(210)      \\
$\Delta_K$/kHz           &       [0.24234]      &        [0.24234]    &        [0.24234]    &        [0.24234]    &     [0.24234]      \\
$\delta_J$/kHz           &      0.01008(77)     &        0.01100(64)  &       0.01004(106)  &        -0.0185(51)  &   0.01207(227)     \\
$\delta_K$/kHz           &       [0.609088]     &        [0.609088]   &        [0.609088]   &        [0.609088]   &    [0.609088]      \\
$\chi_{aa}$ /MHz         &      -4.2651(191)    &       -4.2372( 78)  &       -4.2451(129)  &      -4.2342(193)   &   -4.2360(225)     \\
$\chi_{bb}$ /MHz         &      2.3074(169)     &        2.2880( 66)  &        2.3097(113)  &       2.2833(153)   &    2.3375(208)     \\
$\sigma_{rms}$/kHz       &         30.9         &          15.4       &          10.8       &           15.1      &       1.4          \\
\hline
\end{tabular}
\tablefoot{Numbers in parentheses are the uncertainties in units of the last significant digits. Values in brackets were fixed to those reported by \cite{Zdanovskaia2018}.}
\end{table*}

\subsection{Synthesis of deuterated indene} \label{sec:synthesis}

We synthesized 1d-indene through quantitative lithiation of 1H-indene with n-BuLi, followed by an equally efficient D$_2$O quench to generate the corresponding indenyl-d species. Although the isotopic exchange is essentially quantitative, the final isolated yield (67\,\%) reflects partial loss of material during solvent removal under vacuum because of the volatility of the product. More details are given in Appendix\,\ref{app:synthesis}.

\subsection{Measurements of rotational spectra of D and $^{13}$C indene} \label{sec:experimental}

\begin{figure}
\centering
\includegraphics[angle=0,width=\columnwidth]{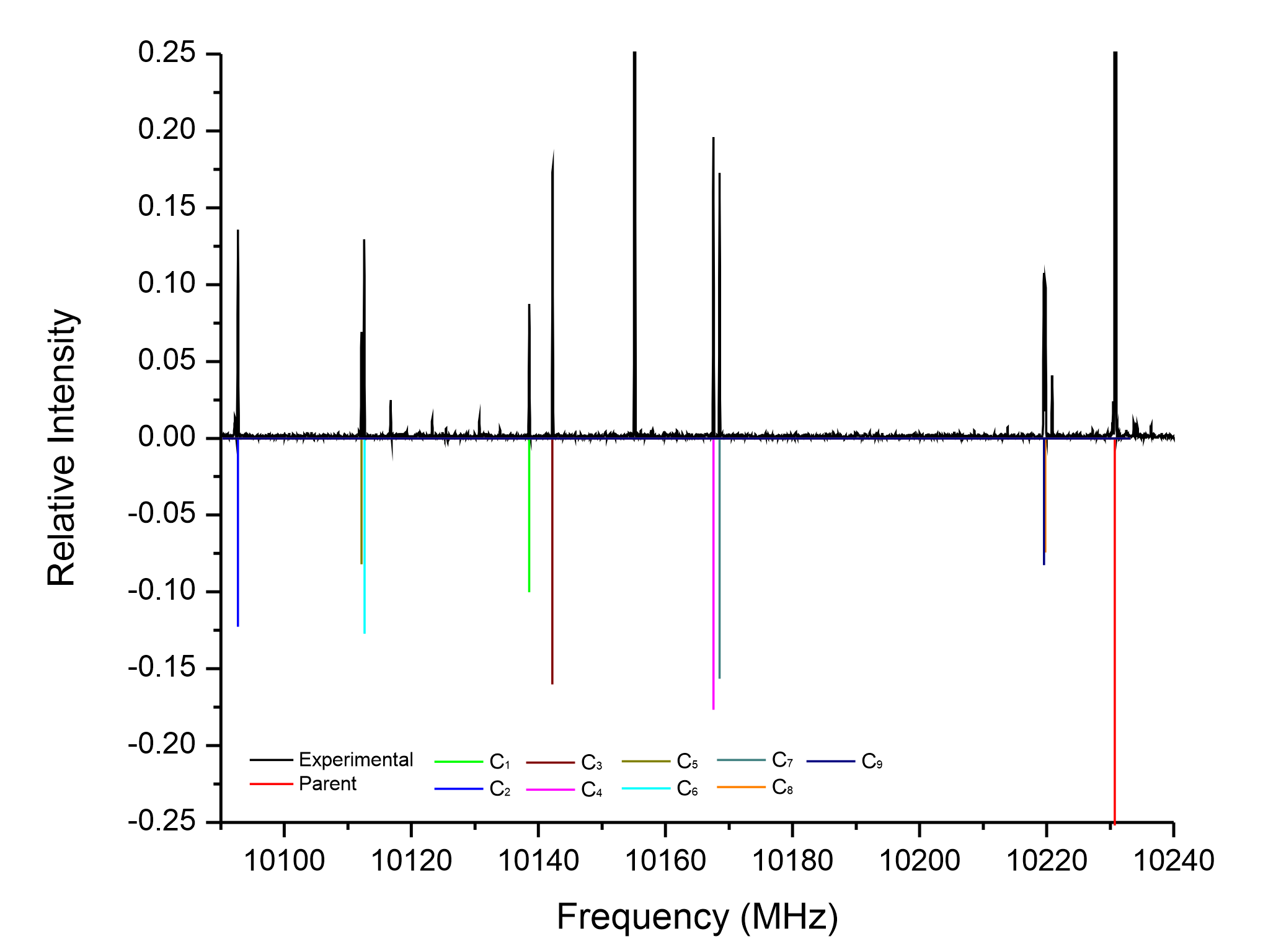}
\caption{Portion of the broadband spectrum recorded between 2-18 GHz for indene, showing the 4$_{0,4}$-3$_{0,3}$ rotational transition of the parent and the nine different $^{13}$C species. The experimental spectrum is shown as the positive black trace. The negative colored traces show simulations at 2 K using the fitted rotational parameters from Table\,\ref{13c_ctes}.} \label{fig:mw}
\end{figure}

In the microwave range (2-18 GHz), we recorded the spectra of 1d-indene and $^{13}$C indene using a broadband chirped-pulse Fourier-transform microwave (CP-FTMW) spectrometer at the Universidad de Valladolid. The instrument is based on the original design of \cite{Brown2008} (see also \citealt{Perez2013}), with adaptations that enable direct broadband excitation and detection across the full frequency window. We placed the sample, either 1d-indene or indene, in a heatable reservoir positioned close to the nozzle orifice and heated it to 80\,\textdegree C to achieve adequate vapor pressure. We entrained the resulting vapor into a pulsed supersonic jet by mixing it with neon at a backing pressure of 3 bar, producing rotationally cold molecular ensembles with effective temperatures near 2 K that greatly simplify the spectral complexity. We polarized the molecules through repeated application of 20 chirped microwave pulses per gas injection cycle, each lasting 2 $\mu$s, at an overall measurement repetition rate of 100 Hz (5 Hz nozzle, 20 pulses per shot). The chirped excitation waveforms were generated with an arbitrary waveform generator (Tektronix AWG 70000, 25 GS/s) and subsequently amplified by a solid-state power amplifier (Qorvo QPB0220N, rated output power 54 dBm) before being broadcast into the vacuum chamber through a horn antenna oriented perpendicular to the molecular beam axis. The transient molecular emission following each excitation event was collected through a second horn antenna, amplified by a low-noise microwave preamplifier (Miteq LNAS-55-01001800-22-10P) and digitized using a high-speed oscilloscope (Tektronix DPO 73304DX, 50 GS/s). We co-added a total of 7 million free-induction decays (FIDs) before Fourier transformation to obtain the final frequency-domain spectrum. We optimized the spectral line shapes by applying a Kaiser–Bessel apodization function, which produced typical full widths at half maximum of approximately 100 kHz. We determined transition frequencies with an estimated accuracy better than 10 kHz and phase-locked all frequency components to a rubidium atomic frequency standard to ensure long-term stability throughout data acquisition.

\begin{table}
\small
\caption{Spectroscopic parameters of 1d-indene.}
\label{table:d-indene}
\centering
\begin{tabular}{lc}
\hline
\hline
$A$/MHz               &   3667.5224(20)  \\
$B$/MHz               &   1559.53360(23)     \\
$C$/MHz               &   1105.53982(19)     \\
$\Delta_J$/Hz        &      33.138(64)    \\
$\Delta_{JK}$/Hz     &      48.30(25)     \\
$\Delta_K$/Hz        &      23.6(13)       \\
$\delta_J$/Hz        &     9.803(39)    \\
$\delta_K$/Hz        &     81.09(82)     \\
$N_t$, $N_f$\,$^a$         &      696, 389         \\
$\sigma_{rms}$/kHz    &        28.9          \\
$J_{min}$, $J_{max}$     &        0, 49          \\
$K_{a,min}$, $K_{a,max}$ &        0, 28          \\
\hline
\end{tabular}
\tablefoot{Numbers in parentheses are the uncertainties in units of the last significant digits.\\
The quantities $^a$\,$N_t$ and $N_f$ refer to the number of distinct rotational transitions and unique transition frequencies, respectively, in the fit.}
\end{table}

\begin{table*}[hb!]
\small
\caption{Spectroscopic parameters for all $^{13}$C isotopic species of indene.}
\label{13c_ctes}
\centering
\begin{tabular}{lccccc}
\hline
\hline
Parameter             &  $^{13}$C$_1$  & $^{13}$C$_2$  & $^{13}$C$_3$  & $^{13}$C$_4$  & $^{13}$C$_5$  \\
\hline
$A$/MHz             & 3736.8219(10)$^a$ & 3774.7942(11)  & 3737.4748(10)   & 3719.9235(16)   &  3761.88283(94)  \\
$B$/MHz             &  1567.35552(43)   & 1551.95861(35) & 1568.02428(47)  & 1576.19575(67)  &  1557.94335(39)  \\
$C$/MHz             &  1112.06213(37)   & 1107.58401(33) & 1112.45740(39)  & 1114.99113(47)  &  1109.50668(36)  \\
$N$\,$^a$             &        72         &       75       &       74        &       59        &        76        \\
$\sigma_{rms}$/kHz  &       18.3        &      18.9      &      17.8       &      14.4       &       19.2       \\
$J_{max}$           &         9         &       15       &       10        &       8         &        10        \\
$K_{a,max}$         &         6         &       4        &       4         &       4         &        4         \\
\hline
\hline
                    & $^{13}$C$_6$  & $^{13}$C$_7$  & $^{13}$C$_8$  & $^{13}$C$_9$ & \\
\hline
$A$/MHz             &  3761.6857(10)   &  3720.2954(12)  & 3761.5052(10)  &  3761.1284(12)  & \\
$B$/MHz             &  1558.06111(47)  &  1576.33934(43) & 1580.65453(38) & 1580.66446(43)  & \\
$C$/MHz             &  1109.55094(36)  &  1115.09721(36) & 1120.94483(32) & 1120.91709(36)  & \\
$N$\,$^a$             &        67        &        74       &       67       &       69        & \\
$\sigma_{rms}$/kHz  &       18.9       &       16.0      &      16.2      &      18.3       & \\
$J_{max}$           &        10        &        10       &       10       &       10        & \\
$K_{a,max}$         &        4         &        4        &       3        &        3        & \\
\hline
\end{tabular}
\tablefoot{Numbers in parentheses are the uncertainties in units of the last significant digits. The quartic centrifugal distortion constants were kept fixed to the values determined for the parent species: $\Delta_J$ = 33.8769 Hz, $\Delta_{JK}$ = 48.699 Hz, $\Delta_{K}$ = 252.575 Hz, $\delta_J$ = 10.0838 Hz, and $\delta_K$ = 85.175 Hz (Cabezas et al., in prep.).\\
The quantity $^a$\,$N$ refer to the number of  rotational transitions included in the fit.}
\end{table*}

We recorded the rotational spectrum of deuterated 1d-indene in the W-band (75–110 GHz) using a chirped-pulse Fourier transform millimeter-wave (CP-FT-mmW) spectrometer located at the University of Valladolid \citep{Steber2025}. The instrument is based on the extension of the chirped-pulse Fourier transform technique into the millimeter-wave regime developed at the University of Virginia \citep{Steber2012,Neill2013} and employs a segmented approach to cover the full 35 GHz bandwidth of the W-band. Broadband chirped-pulse excitation is generated using a Keysight M8190A arbitrary waveform generator (AWG, 12 GSa/s, two-channel).  The AWG output, which spans the 1.56–2.28 GHz frequency range, is upconverted within the BrightSpec Microwave Controller (MWC) unit to the 12.5–18.5 GHz region and subsequently multiplied by a factor of six using a solid-state active multiplier chain (AMC; Virginia Diodes, Inc., WR10-AMC) to produce chirped pulses in the 75–110 GHz range. The millimeter-wave source delivers a nominal output power of approximately 20-30 mW. The radiation is directed into the absorption cell via a WR-10 waveguide horn antenna and collimated using an off-axis parabolic mirror. The absorption cell consists of a 75 cm long stainless-steel single-pass vacuum chamber with an inner diameter of 6.5 cm, sealed at both ends with Teflon windows to allow transmission of the millimeter-wave beam. After traversing the cell, the signal is collected by a second off-axis parabolic mirror and directed into the receiver unit, which is based on a heterodyne subharmonic mixer (Virginia Diodes, Inc., WR10-mixAMC-I). The second channel of the AWG simultaneously generates the local oscillator signal required for downconversion. The resulting intermediate frequency (IF) signal is digitized by a Keysight U1084A digitizer operating at 4 GS/s with real-time signal accumulation. We performed measurements at 80\,\textdegree C in a flow-cell configuration, where we placed the sample in a Schlenk tube connected to the cell inlet and maintained a steady flow at a pressure of approximately 30 mTorr. We co-added a total of 500k signal averages. The estimated frequency accuracy of the measured transitions is better than 50 kHz.

\subsection{Analysis of rotational spectra of D and $^{13}$C indene} \label{sec:analysis}

For 1d-indene we observed a total of 389 rotational lines, 27 in the 2-8 GHz range and 362 between 75 and 110 GHz. These 389 observed frequencies correspond to 696 rotational transitions, as several $a$- and $b$-type transitions with the same $J$ and $K_c$ values but different $K_a$ quantum numbers collapse at the same frequency. All frequency transitions were observed as single lines, even though the deuterium nucleus has a spin of $I$ = 1, and its nuclear quadrupole coupling effects may induce hyperfine splittings. However, these hyperfine splittings are smaller than the experimental line broadening, even for transitions with low $J$ quantum numbers. The list of all measured transitions for this species is provided in the Data availability section. We analyzed all the observed transitions of 1d-indene in a combined fit using the SPFIT program \citep{Pickett1991} with the $A$-reduction Watson's Hamiltonian in the $I^r$ representation \citep{Watson1977}. The analysis yielded the spectroscopic parameters listed in Table\,\ref{table:d-indene} and shows that all quartic centrifugal constants were accurately determined. Attempts to determine any sextic centrifugal constants were unsuccessful.

The spectroscopic parameters in Table\,\ref{table:d-indene} allowed us to accurately predict the rotational transition frequencies for 1d-indene in the Q band with uncertainties below 10 kHz for $a$-type and $b$-type transitions with $K_{a}$ $<$\,10. For the intensity predictions, we assumed that 1d-indene has the same dipole moment components as the parent species \citep{Caminati1993}: 0.50 D and 0.37 D for $\mu_a$ and $\mu_b$, respectively.

The experiment in the 2-18 GHz region for indene allowed us to measure 59--72 lines for each of the nine $^{13}$C species of indene, with $J$ and $K_{a}$ quantum numbers in the ranges 0--10 and 0--4, respectively. As shown in Fig.\,\ref{fig:mw}, the rotational transitions of the $^{13}$C species of indene are easily detected. As for 1d-indene, we fitted all lines using the SPFIT program with the $A$ reduction of Watson's Hamiltonian in the $I^r$ representation. We determined only the rotational constants and fixed the quartic centrifugal distortion constants to the values obtained for the parent species (see the caption in Table \ref{13c_ctes}). Fits in which the centrifugal distortion constants of each $^{13}$C species are fixed to theoretical values, scaled by experimental-to-theoretical correction factors determined for the parent species yield practically identical results. Our quantum chemical calculations show that the predicted distortion constants for each $^{13}$C species differ by 1-3\% from those predicted for the parent species. These changes induce small differences in the Q-band frequency predictions for transitions with $J$\,$<$\,18 and $K_a$\,=\,0-5. For example, for $^{13}$C$_8$, the average difference is 0.9 kHz, with the 45939.943 MHz line showing the largest deviation, 6 kHz. The species with the largest differences is $^{13}$C$_2$, with an average difference of 11 kHz and a maximum deviation of 26.2 kHz for the 48724.57 MHz line. For the remaining $^{13}$C species, the average differences range from 1 to 8 kHz, with maximum values of 17 kHz. We used the spectroscopic parameters of each isotopomer to predict their rotational transition frequencies in the Q band. These predictions have uncertainties of less than 20 kHz for $a$-type and $b$-type transitions with $K_{a}$ values $<$\,4. As for 1d-indene, we adopted the dipole moment components of the parent species for the intensity predictions.

\section{Astronomical search for D and $^{13}$C isotopic species of benzonitrile and indene} \label{sec:search}

We searched for D and $^{13}$C benzonitrile and indene in \mbox{TMC-1} using data from the Q-band line survey carried out with the Yebes\,40m telescope (QUIJOTE; \citealt{Cernicharo2021c,Cernicharo2022}). The data cover the 31.0-50.3 GHz frequency range with a uniform spectral resolution of 38 kHz. We used the latest data, previously described in \cite{Cernicharo2026}, which include observations carried out from November 2019 to July 2024, with a total on-source telescope time of 1509.2 h and a $T_A^*$ rms noise level between 60 $\mu$K at 32 GHz and 180 $\mu$K at 49.5 GHz. We analyzed the data using GILDAS (Grenoble Image and Line Data Analysis Software)\footnote{\texttt{https://www.iram.fr/IRAMFR/GILDAS/}}.

The procedure used to search for D and $^{13}$C benzonitrile and indene in \mbox{TMC-1} consisted of identifying the most favorable lines in the Q band and stacking all of them. We assumed a rotational temperature of 9.0 K for all isotopic species of benzonitrile and indene. This value corresponds to the gas kinetic temperature in \mbox{TMC-1} \citep{Agundez2023b} and to the rotational temperature of the most aromatic cycles for which this parameter could be determined \citep{Cernicharo2021a,Cernicharo2021b}. We chose a reference transition to which we assigned a relative intensity of one and calculated the relative intensities of the rest of the lines assuming local thermodynamic equilibrium (LTE). This procedure allowed us to assign the intensity of the stacked signal to the reference line, from which we determined the column density by adopting a given rotational temperature. We selected all lines with an intensity higher than half that of the most intense predicted line. This intensity threshold provides a good compromise between improved sensitivity and a manageable number of included lines. For the isotopic species of benzonitrile, we excluded lines with $K_a$\,$>$\,2 because their hyperfine splitting is non-negligible compared with the spectral resolution of 38 kHz.

Once we selected the initial set of lines of the target molecule, we extracted \mbox{TMC-1} spectra over $\pm$\,30 km s$^{-1}$ around the expected position of each line. We examined whether a line (or frequency-switching negative artifact) from another molecule was close to the expected position of the target line and, if so, excluded that spectrum from any further analysis. Such spectra cannot be used for line stacking because the target line is contaminated by another line, which would artificially enhance the line stack signal. For the remaining spectra, we cleaned the spectrum by removing the lines (or negative artifacts) from other molecules, so that the noise measured in the stacked spectrum is realistic and is not artificially enhanced by strong positive or negative signals from other molecules present in some of the individual spectra. With the clean set of lines of the target species, we converted the intensity scale to main-beam brightness temperature and scaled each spectrum by the inverse of its predicted intensity relative to the reference transition. We fitted and subtracted a zero-degree polynomial baseline from each spectrum. The (\mbox{TMC-1} spectra are flat because they have been previously reduced, as described in Appendix B of \citealt{Cernicharo2022}). We then stacked the spectra aligned in velocity, using the inverse square of the rms noise level as weight. Although the nominal spectral resolution is uniform in frequency (38 kHz), it is not uniform in velocity across the different individual spectra over the 31.0-50.3 GHz frequency range. The stacking procedure in the CLASS (Continuum and Line Analysis Single-dish Software) program of the GILDAS software produces a stacked spectrum with the poorest velocity resolution of all the individual spectra. This line stack procedure is similar to that used in \cite{Agundez2023a}.

\begin{figure*}
\centering
\includegraphics[angle=0,width=0.95\textwidth]{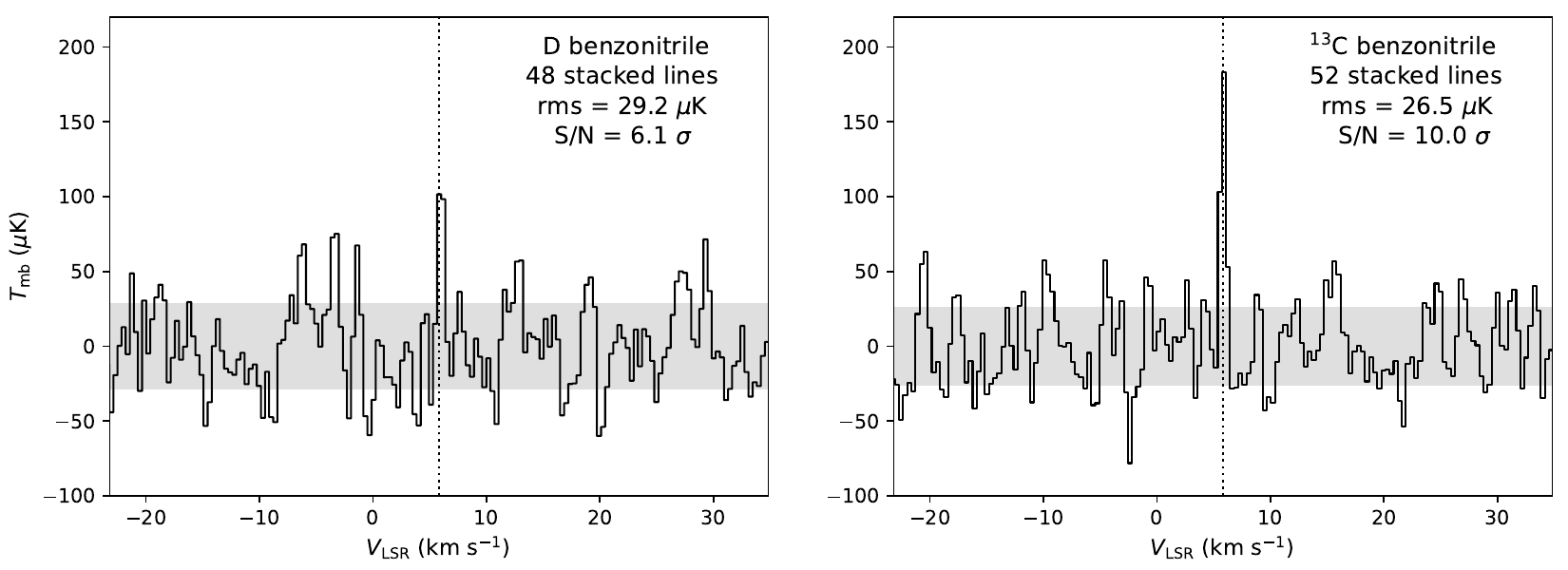}
\caption{Stacked spectra of 48 lines of deuterated benzonitrile (left) and 52 lines of $^{13}$C benzonitrile (right) in \mbox{TMC-1}. The rms noise level at the nominal spectral resolution of each stacked spectrum (0.36 km s$^{-1}$ in both cases) is indicated by a horizontal shadowed band in each panel. The systemic velocity of \mbox{TMC-1} (5.83 km s$^{-1}$) is indicated by a vertical dotted line. Signal centered around the systemic velocity of \mbox{TMC-1} are detected at 6.1 and 10.0\,$\sigma$ for both deuterated and $^{13}$C-substituted benzonitrile, respectively. The detection significance is evaluated using the line area rather than the peak intensity. The signal-to-noise ratio (S/N) is calculated as $\int T_{\rm mb} dv$ / [rms $\times$ $\sqrt{\Delta v \times \delta v}$], where $\int T_{\rm mb} dv$  is the velocity-integrated intensity, rms is the noise level of the stacked spectrum, $\Delta v$ is the full width at half maximum, and $\delta v$ is the spectral resolution in velocity. The adopted reference transition in both cases is 13$_{0,13}$-12$_{0,12}$ (see Sect.\,\ref{sec:search}).} \label{fig:lines_c6h5cn}
\end{figure*}

\begin{figure*}[hb!]
\centering
\includegraphics[angle=0,width=0.95\textwidth]{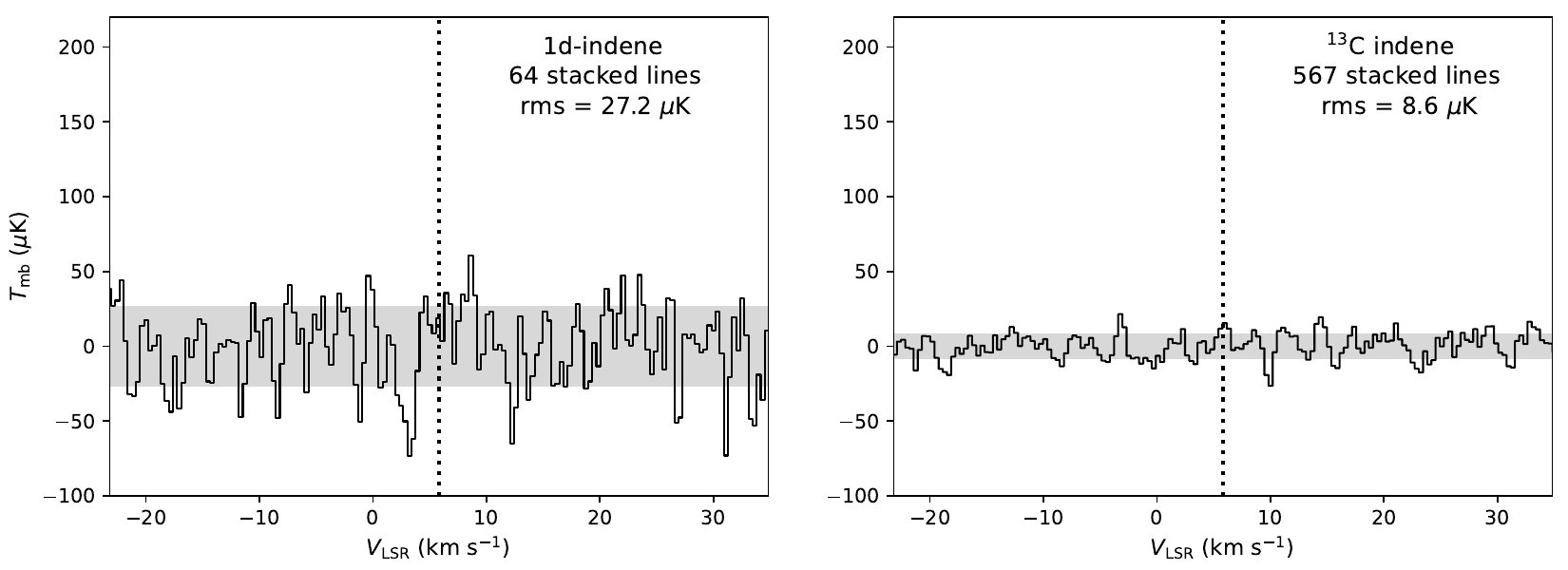}
\caption{Stacked spectra of 64 lines of 1d-indene (left) and 567 lines of $^{13}$C indene (right) in \mbox{TMC-1}. The rms noise level at the nominal spectral resolution of each staked spectrum (0.36 km s$^{-1}$ for 1d-indene and 0.37 km s$^{-1}$ for $^{13}$C indene) is indicated by a horizontal shadowed band  in each panel. The systemic velocity of \mbox{TMC-1} (5.83 km s$^{-1}$) is indicated by a vertical dotted line. The adopted reference transition in both cases is 14$_{0,14}$-13$_{0,13}$ (see Sect.\,\ref{sec:search}).} \label{fig:lines_c9h8}
\end{figure*}

\subsection{D and $^{13}$C benzonitrile}

Deuterated benzonitrile has three isomers, depending on whether deuteration occurs at the ortho, meta, or para position (see Fig.\,\ref{fig:molecules}); the ortho and meta isomers have two equivalent positions for the D atom and para has only one. Therefore, the statistical abundance ratios ortho:meta:para are 2:2:1. We performed the line stack assuming that the three isomers are present according to their statistical relative abundances. Accordingly, we considered only the ortho and meta isomers, which are expected to be twice as abundant as the para isomer. To maximize sensitivity, we stacked the lines of the ortho and meta isomers together, which yielded a total of 48 stacked spectra. The stacked spectrum shows a signal with a signal-to-noise ratio (S/N) of 6.1, centered at 6.01\,$\pm$\,0.06 km s$^{-1}$ (see the left panel of Fig.\,\ref{fig:lines_c6h5cn}), very close to the systemic velocity of \mbox{TMC-1}, 5.83 km s$^{-1}$ \citep{Cernicharo2020}. The shift in the position of the signal with respect to the systemic velocity of \mbox{TMC-1} is likely due to the modest S/N, which leads us to consider the detection tentative. For a rotational temperature of 9.0 K and a circular emission distribution with a radius of 40$''$ \citep{Agundez2023b,Cernicharo2023}, we derive a column density of (1.3\,$\pm$\,0.3)\,$\times$\,10$^{10}$ cm$^{-2}$ for the ortho and meta isomers, and half that value for the para isomer. This value is consistent with the upper limit of 3.0\,$\times$\,10$^{10}$ cm$^{-2}$ derived by \cite{Steber2025} for the ortho, meta, and para isomers.

The $^{13}$C-substituted benzonitrile has five isomers, but as for the deuterated version, we focus on the isomers substituted at positions 2 (ortho) and 3 (meta), since they are statistically twice as abundant as the other $^{13}$C isotopomers. As for deuterated benzonitrile, we assumed that the different $^{13}$C isomers are present with statistical relative abundances and stacked lines of the isomers substituted at positions 2 and 3 together. The spectrum obtained by stacking the 52 spectra shows a 10.0\,$\sigma$ signal centered at 5.85\,$\pm$\,0.05 km s$^{-1}$, in very good agreement with the systemic velocity of \mbox{TMC-1} (see the right panel in Fig.\,\ref{fig:lines_c6h5cn}). For the same rotational temperature (9.0 K) and spatial distribution (radius of 40$''$) as for the deuterated form, we derive a column density of (2.0\,$\pm$\,0.4)\,$\times$\,10$^{10}$ cm$^{-2}$ for the isotopomers substituted at positions 2 and 3, and half that value for the other $^{13}$C isotopomers.

\subsection{D and $^{13}$C indene}

For 1d-indene, we stacked a total of 64 spectra. These spectra correspond to $a$-type transitions with $J_{\rm up}$\,=\,12-21 and $K_a$\,=\,0, 1, 2, 3, which are predicted to be more intense and therefore contribute most to the line stack, and to $a$-type transitions with $J_{\rm up}$\,=\,12-17 and $K_a$\,=\,4, 5, 6, as well as a few $b$-type transitions with $J_{\rm up}$\,=\,6-7, which are weaker and thus contribute less. The stacked spectrum (see the left panel in Fig.\,\ref{fig:lines_c9h8}) shows no evidence of a signal at the systemic velocity of \mbox{TMC-1}. The rms noise level at the nominal spectral resolution of the stacked spectrum, 0.36 km s$^{-1}$, is 27.2 $\mu$K. As for D and $^{13}$C benzonitrile, we adopted a rotational temperature of 9.0 K, a circular emission distribution with a radius of 40$''$, and a line width of 0.60 km s$^{-1}$ \citep{Agundez2023b}. We derived a 5\,$\sigma$ upper limit to the column density of 1d-indene of 1.1\,$\times$\,10$^{12}$ cm$^{-2}$. Given the additional uncertainty associated with the line stack method, compared with the more traditional line-by-line method, we adopted a 5\,$\sigma$ criterion to estimate the upper limit, rather than the more usual 3\,$\sigma$ criterion.

For $^{13}$C indene, the situation is similar to that of 1d-indene, but there are nine different isotopomers. The nine isomers have equal statistical relative abundances; therefore, to maximize sensitivity, we proceeded as for D- and $^{13}$C-benzonitrile and stacked the lines of all isotopomers together. The selected spectra correspond to the same type of transitions considered for 1d-indene, with the $a$-type $K_a$\,=\,0, 1, 2, 3 lines contributing the most. After removing those in which the position of the target line is contaminated, we retained 60-66 spectra for each isotopomer, for a total of 567 spectra (including the nine isotopomers). The larger number of stacked spectra, compared with those for D and $^{13}$C benzonitrile and 1d-indene, results in a significantly lower rms noise level of 8.6 $\mu$K at the nominal spectral resolution of the stacked spectrum, 0.37 km s$^{-1}$ (see the right panel in Fig.\,\ref{fig:lines_c9h8}). This is consistent with the expected dependence of the rms on the inverse square root of the integration time, according to the radiometer equation. That is, the factor-of-three gain in sensitivity from 1d-indene to $^{13}$C indene is simply due to the fact that there are $\sim$\,3$^2$ times more spectra. As for 1d-indene, for a rotational temperature of 9.0 K, an emission distribution with a radius of 40$''$, and a line width of 0.60 km s$^{-1}$, we derive a 5\,$\sigma$ upper limit to the column density of $^{13}$C indene of 3.4\,$\times$\,10$^{11}$ cm$^{-2}$.

\section{Discussion} \label{sec:discussion}

\subsection{D/H ratios}

In general, PAHs contain carbon atoms with aromatic or aliphatic character. The vibrational bands associated with the C-H stretching modes occur at different wavelengths depending on whether the carbon atom has aromatic (at 3.3 $\mu$m) or aliphatic (at 3.4 $\mu$m) character. Similarly, when hydrogen is substituted by deuterium, the vibrational band associated with the C-D stretching mode lies at 4.35 $\mu$m when the carbon atom is aromatic and at 4.65 $\mu$m when it is aliphatic \citep{Hudgins2004}. This allowed us to quantify the D/H ratio at aromatic (D$_{\rm aro}$/H$_{\rm aro}$) and aliphatic (D$_{\rm ali}$/H$_{\rm ali}$) positions in the carriers of AIBs. Studies have inferred D$_{\rm aro}$/H$_{\rm aro}$ ratios of $<$\,0.5\,\% to $<$\,14\,\% in various environments from the upper limits on the 4.35 $\mu$m to 3.3 $\mu$m band intensity ratio\citep{Boersma2023,Draine2025,Yang2025,Steber2025}; it is worth noting that D$_{\rm aro}$/H$_{\rm aro}$ ratios are upper limits in all cases. The D$_{\rm ali}$/H$_{\rm ali}$ ratios determined from the 4.65 $\mu$m to 3.4 $\mu$m band intensity ratio  range from 4.1 to 31\,\% \citep{Boersma2023,Peeters2024,Draine2025,Steber2025}. Therefore, aromatic D/H ratios are expected to be relatively low, while aliphatic D/H ratios are higher. That is, deuterium enrichment tends to occur at aliphatic rather than aromatic sites. The reason for this behavior is not well understood (see discussion in Sect.\,4 of \citealt{Peeters2004}).

Although benzonitrile is not strictly a PAH because it has only one ring, we can use it as a proxy for PAH-like molecules in \mbox{TMC-1}. Specifically, the column density of deuterated benzonitrile determined in \mbox{TMC-1} can be used to estimate the aromatic D/H ratio, while the upper limit on the column density of 1d-indene can be used to derive an aliphatic D/H ratio because the carbon position at which D is located in 1d-indene is aliphatic (position 1 in Fig.\,\ref{fig:molecules}). We define the aromatic D/H ratio as the total number of aromatic C-D bonds divided by the total number of aromatic C-H bonds; similarly, the aliphatic D/H ratio is the total number of aliphatic C-D bonds divided by the total number of aliphatic C-H bonds. We determined the column densities of ortho, meta, and para deuterated benzonitrile in \mbox{TMC-1} to be 1.3\,$\times$\,10$^{10}$ cm$^{-2}$, 1.3\,$\times$\,10$^{10}$ cm$^{-2}$, and 6.5\,$\times$\,10$^{9}$ cm$^{-2}$, respectively, and the column density of parent benzonitrile is 1.2\,$\times$\,10$^{12}$ cm$^{-2}$ \citep{Cernicharo2021b}. From these values, we derive an aromatic D/H of 0.53\,\%. For indene, the column density of the parent, non-substituted species in \mbox{TMC-1} is 1.6\,$\times$\,10$^{13}$ cm$^{-2}$ \citep{Cernicharo2021a} and we determine that of 1d-indene to be $<$\,1.1\,$\times$\,10$^{12}$ cm$^{-2}$. We therefore estimate an aliphatic D/H ratio of $<$\,3.3\,\%. Given the inherent uncertainties associated with the line stack procedure, we estimate the error of the D/H ratios to be 30\,\%.

We can compare the D$_{\rm aro}$/H$_{\rm aro}$ and D$_{\rm ali}$/H$_{\rm ali}$ ratios determined in \mbox{TMC-1} with those measured from AIBs to evaluate whether there could be a chemical link between them. In the case of the D$_{\rm aro}$/H$_{\rm aro}$ ratio, it is difficult to draw definitive conclusions from the value of 0.53\,\% determined in \mbox{TMC-1} because the values determined from AIBs are upper limits ranging from $<$\,0.5\,\% to $<$\,14\,\% \citep{Boersma2023,Draine2025,Yang2025,Steber2025}. However, the D$_{\rm ali}$/H$_{\rm ali}$ provides more meaningful information. The upper limit of $<$\,3.3\,\% derived for \mbox{TMC-1} is of the same order as the lowest value measured from AIBs in the inner regions of the Orion Bar (4.1\,\%;\citep{Peeters2024}). However, the ratios derived in the outer rim of the Orion Bar PDR are considerably higher (15.8-22.5\,\%; \citealt{Peeters2024}), as are those derived in other Galactic and extragalactic environments (12-31\%; \citealt{Boersma2023,Yang2023,Pereira-Santaella2024,Steber2025,Draine2025}).

Therefore, deuteration of the aliphatic carbon of indene is not particularly favored in \mbox{TMC-1}, as is the case for large PAHs in UV-illuminated regions. This can be interpreted as evidence against a top-down origin for PAH-like molecules in cold dark clouds. However, we must be cautious when following this line of reasoning because it is uncertain whether deuterium enrichment holds for AIB carriers in all types of astronomical environment, including diffuse clouds, which is relevant here. \cite{Peeters2024} find that the aliphatic D/H ratio increases by a factor of four when moving from the inner regions to the irradiated rim in the Orion Bar, which could indicate that PAHs are enriched in deuterium in situ in strongly UV-illuminated regions. If that is the case, large PAHs in diffuse clouds may not be rich in deuterium, and therefore our conclusion that PAH-like species in \mbox{TMC-1} are not formed via a top-down mechanism would not be valid. However, it is unclear whether the observed gradient in the band intensity ratio between the inner and outer regions is due to a variation in the abundance of deuterated PAHs or to a variation in their excitation conditions. To address this question properly, observations and models are needed to better understand how large deuterated PAHs are distributed among different interstellar environments.

The D/H ratios derived for benzonitrile and indene, 0.53\,\% and $<$\,3.3\,\%, respectively, are consistent with the range of D/H ratios determined for other molecules in \mbox{TMC-1}, 0.06-3.3\,\% \citep{Cabezas2021a,Cabezas2021b,Cabezas2022,Tercero2024,Cernicharo2024b}. Deuterated molecules detected in \mbox{TMC-1} are typically aliphatic hydrocarbons and N-bearing species. For these molecules, deuteration is usually explained in terms of isotopic fractionation at low temperatures \citep{Roberts2004,Roueff2005}; therefore, isotopic fractionation likely explains the deuterated benzonitrile detected in \mbox{TMC-1}.

\subsection{$^{12}$C/$^{13}$C ratios}

The $^{12}$C/$^{13}$C ratio also provides information on a possible link between the relatively small PAHs present in cold dark clouds and the large PAH-like molecules probed by AIBs. Analyses of micro-sized grains present in meteorites \citep{Amari1990,Bernatowicz1991} show $^{12}$C/$^{13}$C ratios ranging from 0.06 to 16 times the Solar System value of 90 \citep{Anders1989}. This indicates that the grains have an extrasolar origin, most likely associated with material ejected by evolved stars. Interstellar dust grains and the carriers of AIBs are similar in that both are thought to originate in the ejecta of dying stars. Therefore, AIBs carriers should show $^{12}$C/$^{13}$C ratios in line with the values found around evolved stars. However, this has not been confirmed or refuted because it is difficult to constrain the $^{12}$C/$^{13}$C ratio using AIBs. Following this line of reasoning, if the PAH-like molecules detected in \mbox{TMC-1} arise from the fragmentation of inherited AIBs carriers in a top-down scenario, they should also reflect the isotopic composition of evolved stars, which tend to be significantly enriched in $^{13}$C compared to the interstellar medium. For example, \cite{Milam2009} found average $^{12}$C/$^{13}$C ratios of 46 and 18.5 toward C-rich and O-rich AGB stars, respectively, while \cite{Ramstedt2014} derived average values of 34 (C-rich) and 13 (O-rich), although with a significant scatter. Planetary nebulae tend to be even richer in $^{13}$C, with $^{12}$C/$^{13}$C ratios in the range 1.0-13.2 \citep{Ziurys2020}.

From the column density of $^{13}$C-substituted benzonitrile (2.0\,$\times$\,10$^{10}$ cm$^{-2}$ for positions 2 and 3 and half that value for the other positions) and the upper limit to the column density of $^{13}$C-indene ($<$\,3.4\,$\times$\,10$^{11}$ cm$^{-2}$), together with the column densities of the corresponding parent species, 1.2\,$\times$\,10$^{12}$ cm$^{-2}$ for benzonitrile \citep{Cernicharo2021b} and 1.6\,$\times$\,10$^{13}$ cm$^{-2}$ for indene \citep{Cernicharo2021a}, we derive $^{12}$C/$^{13}$C ratios of 126 and $>$\,55 for benzonitrile and indene, respectively. As for deuterium, we estimate an uncertainty of 30\,\% for these $^{12}$C/$^{13}$C ratios. The derived values, in particular that of benzonitrile, are too high to align with the low $^{12}$C/$^{13}$C ratios found around evolved stars, which argues against a chemical link between the small PAH-like molecules present in \mbox{TMC-1} and the AIBs carriers formed in the ejecta of evolved stars.

In the local interstellar medium, the $^{12}$C/$^{13}$C ratio has been determined to be 59\,$\pm$\,2 \citep{Lucas1998}, 69\,$\pm$\,6 \citep{Wilson1999}, 68\,$\pm$\,15 \citep{Milam2005}, 70\,$\pm$\,2 \citep{Sheffer2007}, 76\,$\pm$\,2 \citep{Stahl2008}, 74.4\,$\pm$\,7.6 \citep{Ritchey2011}, and 49-71 at the Galactocentric distance of the Sun \citep{Yan2023}. These studies show good agreement, with $^{12}$C/$^{13}$C ratios in the range 59–76. In \mbox{TMC-1}, $^{12}$C/$^{13}$C ratios have been determined from molecules such as HC$_3$N, HCCNC, HNC$_3$, CS, H$_2$CS, and HCS$^+$, and they lie in the range 69-106 \citep{Cernicharo2024b,Tercero2024,Fuentetaja2025}. The lower limit on the $^{12}$C/$^{13}$C ratio of $>$\,55 derived for indene is consistent with the isotopic ratios observed in the interstellar medium, in general, and in \mbox{TMC-1}, in particular. In the case of benzonitrile, the $^{12}$C/$^{13}$C ratio of 126 is somewhat above the typical values found in \mbox{TMC-1}, which are in the range 69-106; this probably reflects the uncertainties associated with the line stack method. In any case, the value is clearly much higher than the typical values found around evolved stars.

In summary, the D/H and $^{12}$C/$^{13}$C ratios determined for benzonitrile and indene in \mbox{TMC-1} are consistent with typical values of cold dense clouds and do not indicate a particular enrichment in either D or $^{13}$C, as could be expected for AIB carriers. This evidence supports the idea that PAH-like molecules in \mbox{TMC-1} do not form through the fragmentation of large PAHs, but are most likely formed in situ through a bottom-up scenario.

\section{Conclusions} \label{sec:conclusions}

We searched for deuterated and $^{13}$C-substituted benzonitrile and indene in \mbox{TMC-1} using data from the QUIJOTE line survey. We gathered spectroscopic data for benzonitrile from previous experiments, while for indene we characterized the rotational spectrum of the isotopomer substituted at the aliphatic position (1d-indene) in the 2-18 GHz and 75-110 GHz  frequency ranges and those of the nine singly $^{13}$C-substituted isotopomers in the 2-18 GHz frequency range. We used a line stack method to tentatively detect deuterated benzonitrile and detect $^{13}$C-substituted benzonitrile, but find no evidence of either deuterated or $^{13}$C-substituted indene. We derive an aromatic D/H ratio of 0.53\,\% from benzonitrile and an aliphatic D/H ratio $<$\,3.3\,\% from indene, and $^{12}$C/$^{13}$C ratios of 126 and $>$\,55 for benzonitrile and indene, respectively. By comparing these ratios with the aliphatic D/H ratios derived for large PAH-like molecules from infrared observations and with the $^{12}$C/$^{13}$C ratios observed in evolved stars, we conclude that the PAHs detected in \mbox{TMC-1} are unlikely to form through a top-down scenario involving the fragmentation of large PAHs inherited from a previous evolutionary stage of diffuse cloud.\\

\noindent \textbf{Data availability}\\

\noindent We provide the measured and analyzed rotational transitions and frequency predictions for 1d-indene and the nine $^{13}$C isotopomers of indene. We also provide the $^1$H and $^{13}$C NMR (Nuclear Magnetic Resonance) spectra and the GC-MS (Gas Chromatography-Mass Spectrometry) chromatogram of 1d-indene described in Appendix\,\ref{app:synthesis}. This material is available at \href{https://zenodo.org/records/21337335}{zenodo}.

\begin{acknowledgements}

We thank Sabrina Lorenz from the Universit\"atsbibliothek Kiel for providing us the list of laboratory measured frequencies of the isotopic species of benzonitrile from the dissertation ''Mikrowellenspektroskopische Untersuchungen am Benzonitrile und am Benzonitrile-Argon-Van-der-Waals-Komplex'' by Ursula Dahmen (shelf mark: TNA 32). We acknowledge funding support from Spanish Ministerio de Ciencia, Innovaci\'on, y Universidades through grants PID2020-117925GA-I00, PID2021-125015NB-I00, PID2022-137980NB-I00, PID2022-139933NB-I00, PID2023-147545NB-I00, PID2023-146667NB-I00, PID2024-158277NB-I00, and PID2025-171315NB-I00. We acknowledge financial support from the Department of Education of the Junta de Castilla y Le\'on and FEDER (Grant CLU-2025-1-01 -- Unidad de Excelencia Instituto Universitario CINQUIMA, Universidad de Valladolid), A.L.S. acknowledges grant RYC2022-037922-I financed by MCIU/AEI and the FSE+, C.P. thanks the ERC for the CoG HydroChiral (Grant Agreement No 101124939), D. H. thanks the Junta the Castilla y Le\'on for a predoctoral contract, and J.R.M. and M.A.C. thank the University of Valladolid for a predoctoral contract. D.P. and J.J. thank the Xunta de Galicia (Centro de Investigaci\'on do Sistema Universitario de Galicia accreditation 2023-2027, ED431G 2023/03) and the European Union (European Regional Development Fund - ERDF) and Mar\'ia de Maeztu Excellence Accreditation CEX2025-001612-M funded by MICIU/AEI /10.13039/501100011033.

\end{acknowledgements}

\clearpage

\begin{appendix}

\onecolumn

\section{Synthesis of 1d-indene} \label{app:synthesis}

\subsection{General methods}

All reactions were carried out under argon using oven-dried glassware. Anhydrous Et$_2$O was taken from a MBraun SPS-800 Solvent Purification System. 1H-indene was purchased from Sigma-Aldrich and was used without further purification. TLC was performed on Merck silica gel 60 F$_{254}$ and chromatograms were visualized with UV light (254 and 360 nm). Column chromatography was performed on Merck silica gel 60 (ASTM 230-400 mesh). $^1$H and $^{13}$C NMR spectra were recorded at 500 and 125 MHz (Varian Mercury-300 instrument). GC-MS experiments were conducted using a HP 5973 INERT series and an Agilent HP-5MS.

\subsection{Experimental procedures and characterization data}

\begin{figure}[h!]
\centering
\includegraphics[angle=0,width=\textwidth]{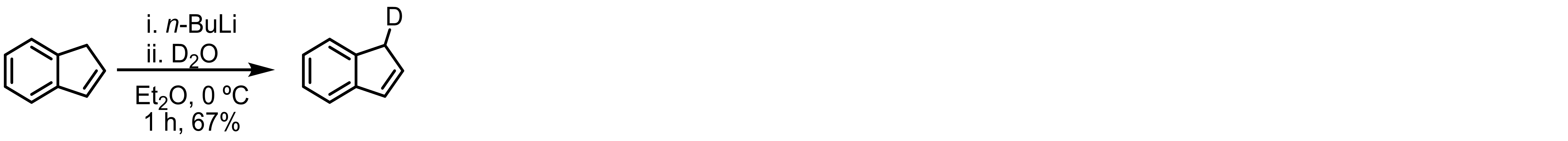}
\caption{General procedure for the synthesis of 1d-indene.} \label{synthesis}
\end{figure}

The synthesis of 1d-indene was done according to an adapted literature procedure \citep{Noland1980}. We added n-BuLi (2.4 M, 18.1 mL, 43.43 mmol) to an stirred solution of 1H-indene (5 g, 43.043 mmol) in dry Et$_2$O (18.3 mL) over 15 minutes at 0\,$^{\circ}$C and stirred for 1 h at room temperature. After that, D$_2$O (2.59 mL, 129.12 mmol) was slowly added at 0\,$^{\circ}$C and the mixture was stirred for 5 minutes at room temperature. Then, the organic layer was separated, washed with NH$_4$Cl (aq), dried over Na$_2$SO$_4$, filtered and concentrated under reduced pressure. The crude compound was purified by column chromatography (hexane, SiO$_2$) yielding 1H-indene-1-d (3.4 g, 67\%) as a yellowish oil.

\textbf{$^1$H-RMN} (500 MHz, CDCl$_3$) $\delta$: 7.50 (d, $J$ = 7.3 Hz, 1H), 7.43 (d, $J$ = 7.5 Hz, 1H), 7.31 – 7.27 (m, 1H), 7.21 (m, 1H), 6.91 (dd, $J$ = 5.6, 2.0 Hz 1H), 6.58 (dd, $J$ = 5.5, 2.0 Hz, 1H), 3.41 – 3.39 (m, 1H) ppm.

\textbf{$^{13}$C-RMN-DEPT} (125 MHz, CDCl$_3$) $\delta$: 145.29 (C), 144.06 (C), 134.54 (CH), 132.55 (CH), 126.65 (CH), 124.95 (CH), 124.13 (CH), 121.36 (CH), 39.32 – 38.86 (t, C$d$) ppm.

\textbf{MS}: GC-MS (EI) m/z (\%): [M]$^+$ observed at 117, consistent with deuterium incorporation.

The $^1$H and $^{13}$ NMR spectra and GC-MS chromatogram of 1d-indene are available (see Data Availability section).

\end{appendix}

\end{document}